\documentstyle[epsf]{mn}
\newif\ifAMStwofonts    
\AMStwofontstrue

\ifoldfss
  \ifCUPmtlplainloaded \else
    \NewTextAlphabet{textbfit} {cmbxti10} {}
    \NewTextAlphabet{textbfss} {cmssbx10} {}
    \NewMathAlphabet{mathbfit} {cmbxti10} {} 
    \NewMathAlphabet{mathbfss} {cmssbx10} {} 
  \fi
  \ifAMStwofonts
    \ifCUPmtlplainloaded \else
      \NewSymbolFont{upmath} {eurm10}
      \NewSymbolFont{AMSa} {msam10}
      \NewMathSymbol{\upi}     {0}{upmath}{19}
      \NewMathSymbol{\umu}     {0}{upmath}{16}
      \NewMathSymbol{\upartial}{0}{upmath}{40}
      \NewMathSymbol{\leqslant}{3}{AMSa}{36}
      \NewMathSymbol{\geqslant}{3}{AMSa}{3E}

    \fi
  \fi
\fi 

\ifnfssone
  \newmathalphabet{\mathit}
  \addtoversion{normal}{\mathit}{cmr}{m}{it}
  \addtoversion{bold}{\mathit}{cmr}{bx}{it}
  \newmathalphabet{\mathbfit} 
  \addtoversion{normal}{\mathbfit}{cmr}{bx}{it}
  \addtoversion{bold}{\mathbfit}{cmr}{bx}{it}
  \newmathalphabet{\mathbfss} 
  \addtoversion{normal}{\mathbfss}{cmss}{bx}{n}
  \addtoversion{bold}{\mathbfss}{cmss}{bx}{n}
  \ifAMStwofonts
    \ifCUPmtlplainloaded \else
      \UseAMStwoboldmath
      \makeatletter
      \new@mathgroup\upmath@group
      \define@mathgroup\mv@normal\upmath@group{eur}{m}{n}
      \define@mathgroup\mv@bold\upmath@group{eur}{b}{n}
      \edef\UPM{\hexnumber\upmath@group}
      \new@mathgroup\amsa@group
      \define@mathgroup\mv@normal\amsa@group{msa}{m}{n}
      \define@mathgroup\mv@bold\amsa@group{msa}{m}{n}
      \edef\AMSa{\hexnumber\amsa@group}
      \makeatother
      \mathchardef\upi="0\UPM19
      \mathchardef\umu="0\UPM16
      \mathchardef\upartial="0\UPM40
      \mathchardef\leqslant="3\AMSa36
      \mathchardef\geqslant="3\AMSa3E
    \fi
  \fi
\fi 

\ifnfsstwo
  \DeclareMathAlphabet{\mathbfit}{OT1}{cmr}{bx}{it}
  \SetMathAlphabet\mathbfit{bold}{OT1}{cmr}{bx}{it}
  \DeclareMathAlphabet{\mathbfss}{OT1}{cmss}{bx}{n}
  \SetMathAlphabet\mathbfss{bold}{OT1}{cmss}{bx}{n}
  \ifAMStwofonts
    \ifCUPmtlplainloaded \else
      \DeclareSymbolFont{UPM}{U}{eur}{m}{n}
      \SetSymbolFont{UPM}{bold}{U}{eur}{b}{n}
      \DeclareSymbolFont{AMSa}{U}{msa}{m}{n}
      \DeclareMathSymbol{\upi}{0}{UPM}{"19}
      \DeclareMathSymbol{\umu}{0}{UPM}{"16}
      \DeclareMathSymbol{\upartial}{0}{UPM}{"40}
      \DeclareMathSymbol{\leqslant}{3}{AMSa}{"36}
      \DeclareMathSymbol{\geqslant}{3}{AMSa}{"3E}
    \fi
  \fi
\fi 

\ifCUPmtlplainloaded \else
  \ifAMStwofonts \else 
    \def\upi{\pi}
    \def\umu{\mu}
    \def\upartial{\partial}
  \fi
\fi

\def\dfrac#1#2{{\displaystyle {#1 \over #2}}}%

\title{Parameters of Core Collapse}
\author[H. Baumgardt et al.]
       {Holger Baumgardt,$^1$ Douglas C. Heggie,$^2$ Piet Hut,$^3$ Junichiro Makino$^1$\\
        $^1$Department of Astronomy, School of Science, The University of Tokyo,
            7-3-1 Hongo, Bunkyo-ku, Tokyo 113-0033, Japan \\
        $^2$School of Mathematics, University of Edinburgh,
         King's Buildings, Edinburgh EH9 3JZ, UK\\
        $^3$Institute for Advanced Study, Princeton, NJ 08540, USA}
\date{Accepted .
      Received ;
      in original form }

\pagerange{\pageref{firstpage}--\pageref{lastpage}}
\pubyear{2002}

\begin{document}

\maketitle

\label{firstpage}

\begin{abstract}
This paper considers the phenomenon of deep
core collapse in collisional stellar systems, with stars of equal mass. \
The collapse takes place on some multiple, $\xi ^{-1},$ of the central
relaxation time, and produces a density profile in which $\rho\propto r^{-\alpha},$
where $\alpha $ is a constant. \ The parameters $\alpha $ and $\xi $
have usually been determined from simplified models, such as gas and
Fokker-Planck models, often with the simplification of isotropy. \ Here we
determine the parameters directly from $N$-body simulations carried out
using the newly completed GRAPE-6.
\end{abstract}


\section{Introduction}

Consider a spherical non-rotating stellar system in dynamic equilibrium. \
Two-body encounters drive a slow evolution of the system. \ In response to
the well known gravothermal instability of such systems (Antonov 1962,
Lynden-Bell \& Wood 1968, Hachisu \& Sugimoto 1978) the core contracts. \
Eventually the central relaxation time is so short that the core loses
thermal contact with the outer parts of the system. \ Thereafter the central
parts of the system evolve in a self-similar manner, unaffected by boundary
conditions (Lynden-Bell \& Eggleton 1980).

In this self-similar regime all central parameters evolve as powers in $\tau
,$ where $\tau $ is the time remaining until collapse ends. Here we neglect 
variations in the Coulomb logarithm in the
expression for the relaxation time. \ If $t_{rc}$ is 
the central relaxation time (defined as in Spitzer 1987, eq.\ 2--62), and $\rho_{c}$
is the central density, it follows that $\dfrac{\dot\rho_{c}}{\rho_{c}}t_{rc}=\xi ,$
where $\xi $ is constant. \ At the same time the density profile
approaches a power law $\rho \propto r^{-\alpha },$ where $\alpha $ is
another constant.

\begin{table*}
\caption{Determinations of $\alpha$ and $\xi$}
\begin{center}
\begin{tabular}{llll}
\hline
Source				&$\alpha$	&$\xi$		& Model\\
\hline
Larson (1970)			&$2.41^1$	&$0.00160^2$	&anisotropic moment	\\
Louis (1990)$^3$                &$2.20$         &$0.00212$      &isotropic moment (eigenvalues)\\
Louis (1990)$^3$                &$2.23$         &$0.00123$      &anisotropic moment (eigenvalues)\\
Lynden-Bell \& Eggleton (1980)  &$2.208$        &$-$            &isotropic gas (eigenvalue)\\
Louis \& Spurzem (1991)$^3$     &$2.23$         &$-$            &anisotropic gas (eigenvalues)\\
Cohn (1980)                     &$2.23$         &$0.0036$       &isotropic Fokker-Planck        \\
Heggie \& Stevenson (1988)      &$2.23$         &$0.00364$      &isotropic Fokker-Planck (eigenvalues)\\
Takahashi (1993)                &$2.23$         &$0.00365^4$    &isotropic Fokker-Planck (eigenvalues)\\
Cohn (1979)                     &$2.27$         &$0.006$        &anisotropic Fokker-Planck \\
Takahashi (1995)                &$2.23$         &$0.0029$       &anisotropic Fokker-Planck\\
Duncan \& Shapiro (1982)	&$2.2$		&$0.006^5$	&Monte Carlo anisotropic Fokker Planck	\\
Joshi et al.\ (2000)            &$2.2$          &$-$            &Monte Carlo anisotropic Fokker Planck  \\
Giersz \& Heggie (1994)         &$2.17^6$       &$-$            &$N = 500$ (average of $\sim50$ cases)\\
Makino (1996)			&$2.36$		&$-$		&$N = 32k^7$\\
This paper 			&$2.26$		&0.0030         &$N = 8k - 64k$\\
\end{tabular}
\end{center}

\begin{flushleft}
Notes  \\

1. From eq.(43), $\rho_c$ 

2. From eq.(44) 

3. Several variants are considered in these papers. The values quoted are those highlighted by the
authors in the abstract

4. Assumes relation for $f$ (distribution function), $\rho_c$ and
central velocity dispersion for
a Maxwellian.

5. Depends on the assumed value of the Coulomb logarithm; see Spitzer (1987, p.95)

6. From Fig.14 

7. From Fig.4 
\end{flushleft}
\end{table*}

Very little can be said about $\alpha $ and $\xi $ on general grounds. \
Since the core is nearly isothermal we may expect that the evolution
timescale is much larger than $t_{rc},$ and so $\xi <<1.$ \ Lynden-Bell \&
Eggleton showed on physical grounds that $2<\alpha <2.5.$ \ It has been
claimed (Lancellotti \& Kiessling 2001) that $\alpha =3,$ on the basis of the
scale invariance of the Fokker-Planck equation. \ It is shown in appendix A
that this is too restrictive a condition.

Precise determination of the parameters $\alpha $ and $\xi$ have been
obtained by a variety of methods (Table 1). 
Lynden-Bell \& Eggleton themselves determined parameters equivalent to $%
\alpha $ and $\xi ,$ using an isotropic gaseous model of a stellar system,
by determining the self-similar solution directly. \ This is an eigenvalue
problem in which their two parameters are eigenvalues. In common with
all gaseous models, the result
for $\xi$ depends on a constant which is usually determined by
comparison with results of some other method, and so the value is not
given in the Table. \ Much earlier,
Larson (1970) determined equivalent parameters by analysing a time-dependent
solution of an anisotropic model based on moments of the Fokker-Planck
equation.

These two methods (eigenvalue problems and analysis of time-dependent solutions) have
been applied by a number of authors using various models and are listed in
Table 1. \ Where necessary their results have been converted to yield values
of $\alpha $ and $\xi $ using relations among core parameters developed by
Lynden-Bell \& Eggleton, except that we uniformly use Spitzer's relaxation
time.  For the theoretical models, the value of the Coulomb logarithm
used in the model is assumed to exactly cancel that in Spitzer's
definition.  In several cases no values were given by the authors themselves, and so we
have added notes to indicate our source for the values given.
Only systems with stars of equal mass are considered here.

While earlier discussions of this topic (e.g. Spitzer 1987, Louis 1990, Louis \& Spurzem
1991) were restricted to comparison of results from simplified models, 
our main aim in this paper is to add data from new $N$-body simulations. \ These
and earlier $N$-body results
are identified in col. 4 of Table 1 by the value of $N$ used, or the range of
$N$. \ 

\bigskip 

\section{Results of $N$-body simulations}

We have performed a new series of $N$-body simulations of isolated clusters starting from
Plummer profiles and containing $N = 8192$ to 65536 stars. The
simulations were carried out on the recently finished GRAPE-6 boards at
Tokyo University, using a specially adapted version of the fully collisional 
$N$-body code NBODY4 (Aarseth 1999). All runs were performed well into the
post-collapse phase. Details of the runs can be found in Table 2.

We first determined the position of the cluster center, using the method of Casertano \& Hut (1985).
According to Spitzer (1987, p.\ 149), the average density $\rho_c$ inside the core radius $r_c$ is 0.517 
times the central density $\rho(0)$ in an isothermal model. Since also the following relation,
connecting 
$r_c$, the 3d-velocity dispersion in the core $\sigma_c$, and the central density, holds for an isothermal model:
\begin{equation}
\sigma_c^2 = \frac{4 \pi}{3} G \rho(0) r_c^2 \;\; , 
\end{equation}
the core radius can be estimated by the following relation: 
\begin{equation}
r_c^2 = 0.517 \frac{3 \; \sigma_c^2}{4 \; \pi G \rho_c}
\label{one}
\end{equation}
Although our clusters initially do not follow isothermal density profiles, 
their cores approach
isothermal profiles as core-collapse proceeds. 
To calculate $r_c$, we first estimated the average density $\rho_c$ inside the core and the
3d-velocity dispersion from the innermost 1\% of the stars and obtained a value for $r_c$ 
from eq.\ \ref{one}.
$\sigma_c$ and $\rho_c$ were then calculated using the stars inside $r_c$ and the whole procedure 
was repeated for 10 iterations. We found that this was enough to determine $r_c$ 
and $\sigma_c$ with sufficient accuracy, 
since after 10 iterations the relative changes of these values between successive iterations 
were of order 1\% or less.
Finally, the core collapse rate $\xi$ was calculated by comparing the core
density at two sufficiently separated points in time and calculating the
central relaxation time at the midpoint.
Fig.\ 1 shows the evolution of $\xi$ as a function of the 
scaled energy $x_0 = - 3 \phi(0)/\sigma_c^2$ for the 64K model.
Here $\phi(0)$ is the central potential, calculated by excluding the
star nearest to the cluster centre. 
\begin{figure}
\epsfxsize=8.3cm
\begin{center}
\epsffile{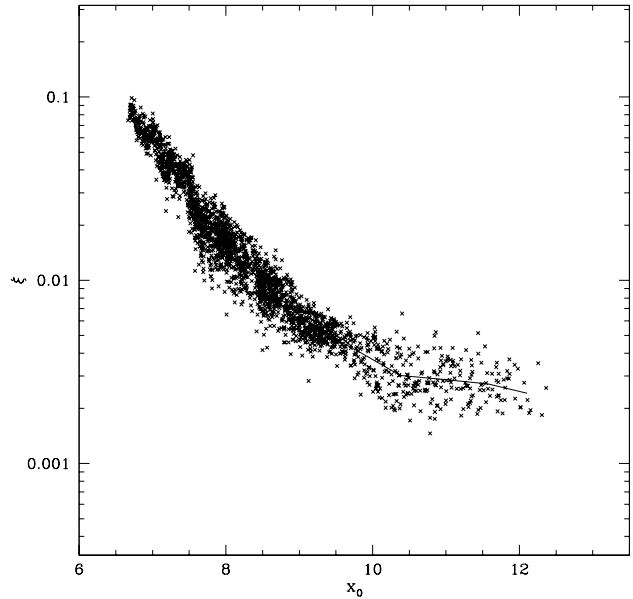}
\caption{Core collapse rate $\xi$ as a function of central escape energy $x_0$
 for the 64K run. Points denote individual values calculated for all times when
 data was stored. The solid line shows the run of the mean $\log(\xi)$. $\xi$ 
 decreases steadily as core collapse proceeds, and becomes constant at around $x_0=10$.}
\end{center}
\label{figxi}
\end{figure}

The $N$-body data is noisy since the time derivative of the central density
is used for calculating $\xi$. Nevertheless, it can clearly be seen that 
$\xi$ decreases until about 
$x_0 = 10$, after which it becomes nearly constant. This is in good agreement
with the behavior 
found by Cohn (1980) and Takahashi (1995) in their Fokker-Planck simulations.
Core-collapse is completed at around $x_0 = 13.0$. Taking the mean $\xi$ from 
all data points with $x_0 > 11$, we obtain a limiting value of $\xi = 0.0029$
for this run. Similar values are obtained for runs with other $N$ (see Table 2).  
Taking the mean over all performed runs, we obtain $\xi = 0.0030$, which is
in good agreement with what Takahashi (1995) obtained from anisotropic Fokker-Planck calculations (see
Table~1).

\begin{table}
\caption{Results for the core-collapse time $T_{CC}$, $\alpha$ and $\xi$ from $N$-body simulations}
\begin{center}
\begin{tabular}{crrrr}
\hline
$N$        &   8192 &  16384 & 32768  & 65536 \\
\hline 
$N_{Sim}$  &      8 &      3 &     2  &      1 \\
$<T_{CC}>$ &   1967 &   3640 &  6796  &  12218 \\
$<\alpha>$ &   2.24 &   2.26 &  2.28  &   2.26 \\
$<\xi>$    & 0.0030 & 0.0031 & 0.0030 & 0.0029 \\
\end{tabular}
\end{center}
\end{table}

In order to measure the density gradient at the time of core-collapse, we proceeded in
the following way: The time of maximum core contraction was determined from the time when 
the potential energy at the cluster centre ($\phi(0)$, calculated as above) reached its 
first minimum. We then calculated
the stellar density as a function of distance from the centre and fitted
power-law distributions to the density profile inside 0.1 half-mass radii.
The slope of the best fitting 
power-law was determined by a KS-test. Mean values of $\alpha$ can be found in Table 2. 
In general, we obtain 
somewhat larger values than the Fokker-Planck calculations, and our results seem to be only
marginally compatible with the $\alpha$ preferred by most Fokker-Planck and gas calculations: 
$\alpha  = 2.23$. For the range of particle numbers studied, no clear change of $\alpha$ 
with $N$ can be seen.

\begin{figure}
\epsfxsize=8.3cm
\begin{center}
\epsffile{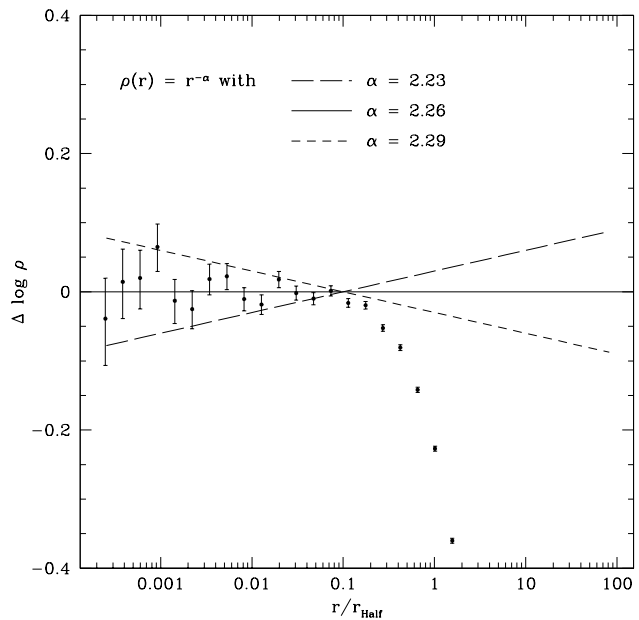}
\caption{Density profile of the combined set of $N$-body runs compared with 3 different
 power-laws $\rho(r) = r^{-\alpha}$. All data is divided by the density of a power-law
 profile with $\rho(r) = r^{-2.26}$. Power-law profiles are adjusted such to contain the same 
 number of stars as the $N$-body data inside $r = 0.1\, r_{Half}$. The best fit is achieved 
 for $\alpha = 2.26$.
 The uncertainty of this value does not seem to be larger than 0.02, ruling out a
 value of $\alpha = 2.23$ in the $10^4 < N < 10^5$ range.}
\end{center}
\label{fig:alpha}
\end{figure}

Fig.\ \ref{fig:alpha} compares the combined $N$-body data from runs with $N = 8192$ to 65536 stars with 
various power-law profiles. In order to determine the slope of the density profile, we fitted data up to a 
maximum radius
of $r = 0.1 \, r_{Half}$. Fig.~\ref{fig:alpha} shows that outside this radius, the density profile cannot be
fitted by a single
power-law any more, while inside from there the value of $\alpha$ will not depend on the maximum
radius used for the fit. Inside $r = 0.1 \, r_{Half}$, we obtain a slope of $\alpha  = 2.26$ for
the density profile.
A value of $\alpha  = 2.23$ would give a bad fit to the combined $N$-body data, but might
be possible in the $N \rightarrow \infty$ limit if $\alpha$ is changing slowly with $N$. 
A value of $\alpha  = 3.0$ is completely ruled out by our $N$-body simulations.

Simplified anisotropic models for star cluster evolution
(e.g. Giersz \& Spurzem 1994) predict that, at a given radius inside
the self-similar regime, the anisotropy approaches a finite value at
the end of core collapse.  Our results for the evolution of the anisotropy
profile are shown in Fig. 3. At the end of core collapse, 
we obtain an anisotropy profile closely resembling Fig.~2 in Giersz \&
Spurzem, though the value of the anisotropy parameter A at small radii
is a little smaller, around 0.2.  The maximum value of about 1.1
(which occurs outside the self-similar regime) is a little larger than
theirs.  The time evolution of the anisotropy within different
Lagrangian shells closely resembles the results from averaged
1000-body models shown in their Fig.~11.
\begin{figure}
\epsfxsize=8.3cm
\begin{center}
\epsffile{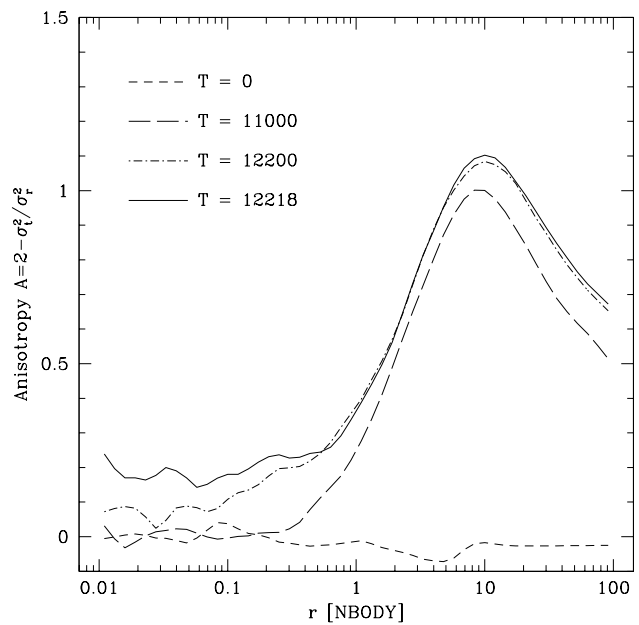}
\caption{Anisotropy profile for the run starting with $N=65536$ stars for 4 different
 times. The anisotropy parameter $A$ rises as core-collapse proceeds until it becomes
 nearly constant inside the self-similar regime.}
\end{center}
\label{fig:aniso}
\end{figure}

\bigskip

\section{Conclusions}

Late core collapse of stellar systems is one of the few regimes where
the evolution becomes relatively simple.  For systems with particles
of equal mass, simple arguments imply that many properties
(central density and velocity dispersion, core radius, density
profile, etc.) approach simple limiting forms, which can be
characterised by just two parameters.  A common choice for these is
the dimensionless rate of increase of the central density, $\xi$, and
the index of the power-law dependence of density on radius outside the
core, $-\alpha$.  

In this paper we explain why $\alpha$ is not determined, as has been
argued, by the scale invariance of the Fokker-Planck equation.  We
review historical determinations of these parameters based on
numerical solutions of the Fokker-Planck equation and other simplified
models for the evolution of stellar systems.  Our main contribution,
however, is to present determinations of these parameters directly
from new large direct $N$-body computations.  We find that the core
collapse rate $\xi$ ($= 0.0030$) agrees satisfactorily (within the
statistical error of the N-body results) with those determined by
better simplified models.  The density profile index $\alpha$,
is slightly steeper, our best value being about $2.26$.

\bigskip    

\section*{Acknowledgments}
We thank the referee for his helpful comments.

\bigskip

\appendix

\section{Scaling of the Fokker-Planck equation}

It has been shown by Lancellotti \& Kiessling (2001) that the Fokker-Planck equation admits a
unique scale invariance, and that therefore the self-similar solution
requires a limiting density profile $\rho \propto r^{-3},$ i.e. $\alpha =3.$ \
Here it is shown why the value of $\alpha $ is not determined by the scale
invariance of the Fokker-Planck equation.

Consider first the model problem
$$
\dfrac{\partial f}{\partial t}+{\mathbf v.}\dfrac{\partial f}{\partial 
{\mathbf r}}-\dfrac{{\mathbf r}}{r^{3}}.\dfrac{\partial f}{\partial {\mathbf v}}%
=f^{2},  
\eqno{(A1)}
$$
which can be interpreted as a Fokker-Planck equation for a distribution of
Keplerian oscillators with energy $E=\dfrac{1}{2}v^{2}-1/r.$ \ The right side
of eq. (A1) is a simple collision term, chosen only so that it has the same
scaling property as in the Fokker-Planck equation of collisional stellar
dynamics. \ Indeed eq. (A1) admits the unique scaling $f\rightarrow \mu f,$ $%
t\rightarrow \mu ^{-1}t,\;\mathbf{r\rightarrow }\mu ^{-2/3}\mathbf{%
r,v\rightarrow }\mu ^{1/3}\mathbf{v.}$ \ It follows that eq. (A1) admits
self-similar solutions of the form 
$$
f=t^{-1}F\left( {\mathbf r}t^{-2/3},{\mathbf v}t^{1/3}\right) ,
\eqno{(A2)}
$$
where $F$ is some function satisfying a certain partial differential
equation.
Hence the space density is $\rho =t^{-2}\int F\left( {\mathbf r} t^{-2/3},%
{\mathbf v}^{\ast }\right) d^{3}{\mathbf v}^{\ast },$ where ${\mathbf v}^{\ast
}={\mathbf v} t^{1/3.}$ \ This is stationary at large $r$ only if $\rho \propto r^{-3},$
i.e. $\alpha =3.$

These are not the {\sl {only}} self-similar solutions of eq. (A1), however.
\ There are also solutions of the form
$$
f=t^{-1}F\left( t^{\beta }E
\right) ,  \eqno{(A3)}
$$
where $\beta $ is any constant and $F$ satisfies a certain ordinary
differential equation. \ Such solutions are certainly self-similar, in the
sense that the function $f$ evolves by time-dependent scalings of $f,\mathbf{%
v}$ and $\mathbf{r.}$ \ The reason why eq. (A1) admits a wider class of
self-similar solutions than those of the form (A2) is that solutions of the
form (A3) also satisfy the differential equation
\[
{\mathbf v.}\dfrac{\partial f}{\partial {\mathbf r}}-\dfrac{{\mathbf r}}{r^{3}}.%
\dfrac{\partial f}{\partial \mathbf{v}}=0
\]
and hence also the simpler Fokker-Planck equation
$$
\dfrac{\partial f}{\partial t}=f^{2}.  \eqno{(A4)}
$$
This pair of differential equations admits a much wider class of scalings
\[
f\rightarrow \mu f,\;t\rightarrow \mu ^{-1}t,\;\mathbf{v}\rightarrow
\nu \mathbf{v,\;r}\rightarrow \nu^{-2}\mathbf{r\;.}
\]

Another interpretation of this situation is to observe that (A4) is the
orbit-averaged version of eq. (A1) (cf. Spitzer 1987). \ In the language of
stellar dynamics it is the equation obeyed by solutions which evolve slowly,
on a timescale $t_{ev}$ much longer than the orbital or crossing timescale, $%
t_{cr}.$
Indeed the first term on the left of eq. (A1), and the term on the right,
are of order $f/t_{ev}$, while the remaining terms of eq. (A1) are of order $%
f/t_{cr}.$ \ If we restrict attention to self-similar solutions of the
form of eq. (A2)
we are, in effect, insisting that $t_{ev}\propto t_{cr}.$ \ \ Indeed the
scaling of eq.(A1) is exactly the same as that of the $N$-body equations
\[
{\mathbf \ddot{r}}_{i}=-G\sum\limits_{j\neq i}m_{{j}}\dfrac{{\mathbf r}_{i}-%
{\mathbf r}_{j}}{\left| {\mathbf r}_{i}-{\mathbf r}_{j}\right| ^{3}},
\]
and the analogues of eq.(A2) are then the familiar homothetic
solutions in which ${\mathbf r}_i\propto t^{2/3}$ (cf. Arnold et al.\ 1997, p.65).

For the Fokker-Planck equation of stellar dynamics the situation is a little
more complicated. \ Its general and orbit-averaged forms are
$$
\dfrac{\partial f}{\partial t}+{\mathbf v}.\dfrac{\partial f}{\partial \mathbf{r}}%
-\nabla \phi .\dfrac{\partial f}{\partial \mathbf{v}}=\left( \dfrac{\partial
f}{\partial t}\right) _{c}  \eqno{(A5)}
$$
and
$$
\dfrac{\partial f}{\partial t}+\dfrac{\partial f}{\partial E}\left\langle 
\dfrac{\partial \phi }{\partial t}\right\rangle =\left\langle \left( \dfrac{%
\partial f}{\partial t}\right) _{c}\right\rangle ,  \eqno{(A6)}
$$
where $\left( \dfrac{\partial f}{\partial t}\right) _{c}$ is the collision
term, $\phi $ is the potential, $E=\phi +\dfrac{1}{2}v^{2}$ and $%
\left\langle {}\right\rangle $ denote an orbit average. \ Eq. (A6) is not
obtained from eq. (A5) by assuming that $f=f\left( E,t\right) .$ \ Instead,
we must assume that a solution of eq. (A5) may be expanded in the form $%
f=f_{0}+\varepsilon f_{1}+...$ where $\varepsilon =t_{cr}/t_{ev}$ and $%
f_{0}=f_{0}\left( E,t\right) .$ \ Then $f_{0}$ obeys eq. (A6), while $f_{1}$
and higher terms obey appropriate linearised forms of eq. (A5).

For self-similar core collapse, $f_{0}$ may be taken to be of the form of
eq. (A3).
There is no reason to suppose that $f_{1}$ enjoys the same self-similar
evolution as $f_{0}.$ \ Thus the solution of eq. (A5) for core collapse is
$nearly$ self-similar (up to terms of order $t_{cr}/t_{ev})$ but is not
confined by the highly restrictive scaling properties of eq. (A5) itself. \
As core collapse comes to an end, with a small number of stars remaining in
the core, $t_{ev}$ decreases and becomes nearly comparable with $t_{cr}$. \
Then new phenomena beyond the Fokker-Planck equation become important, such
as formation of binaries in 3-body encounters.


\bsp
\label{lastpage}

\end{document}